\documentclass[graybox, envcountchap]{svmult}
\usepackage{aas_macros}
\usepackage[T1]{fontenc}

\usepackage{mathptmx}        
\usepackage{amsmath}
\usepackage{amssymb}
\usepackage[utf8]{inputenc}
\usepackage{helvet}          
\usepackage{courier}         
\usepackage{dirtree}

\usepackage{makeidx}        
\usepackage{xcolor}
\usepackage{graphicx}        
\usepackage{subfig}

\usepackage[numbers]{natbib}

\usepackage{multicol}        
\usepackage[bottom]{footmisc}

\usepackage{hyperref}        
\hypersetup{colorlinks=true,urlcolor=blue}

\usepackage[misc]{ifsym}

\makeindex             

\begin{document}


\title{Theoretical Instability Models of Quasi-Periodic Eruptions}
\author{Agnieszka Janiuk, Bożena Czerny, Marzena \' Sniegowska, Pulkit Ojha, Shuang-Liang Li, Shivangi Pandey}

\institute{Agnieszka Janiuk (\Letter) \at
Center for Theoretical Physics, Polish Academy of Sciences,
Al. Lotników 32/46, 02-666 Warsaw, Poland \\
\email{agnes@cft.edu.pl}
\and
Bożena Czerny \at
Center for Theoretical Physics, Polish Academy of Sciences,
Al. Lotników 32/46, 02-666 Warsaw, Poland \\
\email{bcz@cft.edu.pl}
\and
Marzena \'Sniegowska \at
Astronomical Institute, Czech Academy of Sciences, Bo\v{c}n\'{i} II 1401, Prague, 14100, Czech Republic\\
\email{sniegowska@mail.asu.cas.cz}
\and
Pulkit Ojha \at
Center for Theoretical Physics, Polish Academy of Sciences,
Al. Lotników 32/46, 02-666 Warsaw, Poland \\
\email{pojha@cft.edu.pl}
\and
Shuang-Liang Li \at
Shanghai Astronomical Observatory, Chinese Academy of Sciences,
80 Nandan Road, Shanghai 200030, People’s Republic of China \\
\email{lisl@shao.ac.cn}
\and
Shivangi Pandey \at
Center for Theoretical Physics, Polish Academy of Sciences,
Al. Lotników 32/46, 02-666 Warsaw, Poland \\
\email{spandey@cft.edu.pl}
}

%
%
\maketitle
\abstract{
The nature of the Quasi-Periodic Eruptions (QPE) is still not firmly established while the phenomenon probes the innermost part of the accretion flow undergoing dramatic changes. In this chapter we critically review the possibility that the observed rapid variability is caused by the accretion disk instability. 
We summarise the defining observational properties of QPE sources: time scales of recurrence, the amplitudes of soft X-ray flares, high coherence over many cycles, and comparatively modest time-averaged luminosities. We then translate these into constraints on the characteristic radius, accretion rate, and energy reservoir of the inner disk. We review the main classes of disk instabilities relevant to supermassive black holes: radiation pressure induced thermal–viscous limit cycles, magnetically mediated inner disk transitions, and coupling to disk winds that can produce the oscillatory behavior. For each scenario we discuss the required parameter space (black-hole mass, Eddington ratio, viscosity prescription, and truncation radius), the predicted spectral and timing signatures, and the degree of fine tuning needed to reproduce the observed duty cycles and flare morphologies. We highlight key tensions, including the extremely soft spectra, short recurrence times compared with standard viscous times at tens of gravitational radii, and the weak or absent contemporaneous optical/UV variability in several systems. {Finally, we outline observational tests that may help distinguish between instability driven and geometrical or orbital models: phase resolved spectroscopy to track the temperature evolution, searches for hysteresis in hardness–intensity diagrams, reverberation and absorption variability tied to winds, and long baseline monitoring to measure drift in period and energy output. }
These diagnostics will clarify whether QPE reflect transitions between accretion disk states driven by known instabilities, or instead require an external trigger.
}


\section{Introduction}

Repeating Extraterrestrial Nuclear Transient (RENT) is a fascinating new window into the dynamical processes taking place in the vicinity of the supermassive black holes hosted by the active and non-active (or rather weakly-active) galaxies. Observing the time evolution we can get insight into the processes taking place even without the spatial resolution of the region where the phenomenon takes place, and such a direct mapping is not within the reach for the coming decades. However, the collected report of the time evolution and the excellent future observational prospects for detecting many more of such events is a motivation to summarize current understanding of these new phenomena. 

The RENT class is by no means homogeneous. At present three separate classes of events are broadly discussed: Tidal Disruption Events (TDE), Quasi-Periodic Oscillations (QPO), and Quasi-Periodic Eruptions (QPE). These classes are not necessarily unrelated: several QPE sources show evidence for a preceding TDE-like episode, suggesting that in at least some systems the QPE phenomenon may emerge during the subsequent evolution of a newly formed or strongly perturbed accretion flow.  QPE may - or may not - share the common underlying mechanism with a special class of Active Galactic Nuclei (AGN) which exhibit Changing Look (CL)/Changing State (CS) behaviour in timescales of years. In the current book we basically limit ourselves to the faster changes, from hours to days/months.

The book consists of several chapters devoted to these topics. The current Chapter is devoted to the \textit{Theoretical Instability Models of QPE.} There are two leading explanations of the QPE: either a TDE event which leads to partial disruption of the star, with the material repeatedly crossing either the accretion disk or its own debris, or it is related to the instabilities in the innermost flow of the accretion disk, either pre-existent or formed in the long TDE event \citet{quintin2025}. By this we mean that an accretion disk around a black hole pre-exists or forms in the process, and the observed variability is related to some instability in the accretion disk flow. A dominant group of models relate the QPE to the orbital motion of the second body (e.g. remnants of the disrupted star) around the supermassive black hole which can periodically deposit the new material, or can repeatedly cross the disk perturbing it. Such models were discussed in Chapter 7A. Our Chapter, 7B, concentrates on intrinsic disk instabilities.

Below, we only shortly summarize the observational background since the observations of QPE will be discussed in details in Chapter 4. Here, it serves as a testbed of the models under discussion.

\section{Observational background}
\label{sec:obs}

In this section, we focus only on those observational aspects which are essential from the point of view of the accretion disk instability as a potential underlying physical mechanism.

The QPE phenomenon has been discovered by \citet{miniutti2019} in the source GSN 069, and this object remains one of the basic sources of information due to the well known history of the source and the data quality.  GSN 069 is a clear example of long Tidal Disruption Event (TDE). It was first detected in 2010, in X-ray band when it unexpectedly brightened by a factor of more than 250 with respect to the upper limit derived 16 years earlier. Yet in 2010 and in 2014 the source did not show any special variability, and QPE phenomenon only appeared in 2018, when the source dimmed considerably. With subsequent systematic dimming the source pulsations turned off and it reappeared in 2023 \citep{minitti_revival2023} when the source became temporarily brighter again. 

The QPE period in GSN 069 was extremely short, only 9 hours, and fairly regular. As the sample has grown, QPE events have also been reported with recurrence times approaching days (e.g. the optically selected TDE AT2019qiz with a $\sim 48$~hrs mean recurrence \citep{nicholl2024}; and AT2022upj \citep{Chakraborty2025}, with a wide spread $\sim 0.5 - 3.5$~days, indicating that the phenomenology spans at least hours to several days. Also, objects with shorter periods were found. Until now (October 2026) we have 13 sources with detected QPE (see e.g. \cite{Jiang2025,2026Baldini}, \cite{2026arXiv260715359W}) which already forms an attractive sample for attempt to search for a common physical mechanism. It is also important to notice that the earliest example of the QPE is in the source RXJ 1301.9+2747\citep[see][]{middleton2015,shu2017} but it was not recognized as QPE until \citet{giustini2020}. 

\subsection{timescales}

Observed QPE recurrence times span for most of the sources
from 2–3 h up to 40–50 h, with the recently found outlier, ZTF19acnskyy/'Ansky' \citep{2025hernandez}
with recurrence times of approximately 10 days. Typical durations of individual eruptions are from less than 1 hour to a few hours,
so the duty cycle is relatively short. The outbursts are not sinusoidal but of a
flare type, with the highest amplitude observed is soft X-ray band, 0.3 - 2 keV, which is often
higher than a factor 10 with respect to the quiescence state. The recent observations of 'Ansky' have shown a period of flares increasing with time and made this source even more challenging to explain from theoretical point of view \citep{2026Chakraborty}.

\subsection{energy constraints of QPE component}

The peak bolometric luminosity of QPE component in GSN 069 is
$3.3 \times 10^{42}$ erg s$^{-1}$ \citep{miniutti2023}, and the total
energy integrated over the QPE period is $6.5 \times 10^{45}$ erg.
Similar values of the peak luminosity were obtained for RX J1301.9+2747
($2.2 \times 10^{42}$ erg s$^{-1}$; \citealt{giustini2024}), and overall
the dispersion in both the peak luminosity and the total luminosity is
not large, with the outburst duration roughly being a fraction of the
recurrence time (see Fig. 13 in \cite{arcodia2024}). Thus the few sources
for which the bolometric QPE excess has been estimated suggest peak
luminosities of a few $\times 10^{42}$ erg s$^{-1}$ and radiated energies of order a few $\times 10^{45}$ erg per eruption.

\subsection{black hole mass and the Eddington ratio}
Current mass estimates place QPE hosts preferentially in the low-mass SMBH regime. In GSN 069, the black hole mass is reported as 
$4 \times 10^{5} M_{\odot}$. RX J1301.9+2747 has a substantially higher inferred mass, $\sim (0.8- 2.8) \times 10^{6} M_{\odot}$. 
While individual objects remain uncertain at the factor of a few level (depending on method), the broader empirical fact is that QPEs are
mostly associated with $M \le 10^{7} M_{\odot}$ black holes \citep{miniutti2019}.
Disk-instability scenarios are particularly sensitive to the Eddington ratio, $\lambda=L/L_{\rm Edd}$. 

 However, observational constraints of $\lambda$ are difficult as they relay both on the black hole mass estimate as well as some bolometric corrections when the observations do not cover well the whole spectrum. Fixed bolometric correction to a soft X-ray flux likely overestimate the Eddington ratio while using only black body fit to QPE component can underestimate it considerably.

 For RX J1301.9+2747, the accretion rate has been estimated at $\dot m \sim 0.14$ (in Eddington units). In GSN 069, the 2023 reappearance of QPEs indicates that eruptions are preferentially present below a quiescent luminosity threshold corresponding to $\lambda\sim0.4$ (for 
$10^{6} M_{\odot}$ BH), suggesting that detectability and/or the underlying limit cycle depends sharply on the background disk state \citep{giustini2020}.

Below, in Table \ref{table:data} we summarize the known estimates for black hole mass and Eddington ratio
for a handful of known sources.

\begin{table}
\caption{Black hole masses and estimates of the Eddington ratios in active phase for QPE sources}
\label{table:data}
\begin{tabular}{p{2cm}p{2.2cm}p{2.4cm}p{2cm}p{4.9cm}}
\hline\noalign{\smallskip}
Source & model & $M_{BH}$ & $\lambda_{Edd} $ & reference\\
& comment & $M_{\odot}$\\
\noalign{\smallskip}\svhline\noalign{\smallskip}
GSN 069 &  & $4 \times 10^5$ & $\sim 0.46$ & \citet{miniutti2019} \\
GSN 069 & high spin & $7.5^{+1.0}_{-1.5} \times 10^6$ & $0.035\pm 0.007$  & \citet{Guolo_fitting_2025}\\
GSN 069 & low spin & $1.5^{+0.5}_{-0.3} \times 10^6$ & $0.148\pm 0.037$  & \citet{Guolo_fitting_2025}\\
RX J1301 & active 2000–2022 & $(0.8-2.8) \times 10^6 $ & 0.14 & \citet{dewangan2000,giustini2020,shu2017}\\
eRO-QPE1 &  & $(0.2 - 2.1) \times 10^6$ & $\sim 1$ & \citet{arcodia2021} \\
eRO-QPE2 &  &  $10^{5.9 \pm 0.3}$ & $0.13^{+0.18}_{-0.07}$ & \citet{weavers2025}\\
eRO-QPE3 &  & $(0.4-1.7) \times 10^6$ & $\sim 0.0013$ & \citet{arcodia2024} \\
eRO-QPE4 &  &  $(17-68) \times 10^6$ & $\sim 0.0064$ & \citet{arcodia2024} \\
AT2019qiz &  &  $10^{6.3^{+0.3}_{-0.2}}$ & $0.04 \pm 0.01$ & \citet{nicholl2024}\\
AT2022upj & & $ 5.6^{+3.3}_{-2.5} \times 10^6$ & 0.002 - 0.014 & \citet{Chakraborty2025}\\
ZTF19acnskyy&  & $(6.09 \pm 0.55) \times 10^6$ & $\sim 0.1$ & \citet{Guolo2024}\\
\noalign{\smallskip}\hline\noalign{\smallskip}

\end{tabular}

\medskip
\parbox{\linewidth}{\footnotesize
$^{a}$ eRO-QPE2 is likely a TDE remnant with an outer disk radius of
$R_{\rm out} = 343^{+202}_{-138}\,R_g$.
Some sources are confirmed TDEs, including GSN 069
(\citealt{miniutti2019}), AT2019qiz (\citealt{nicholl2024}),
and AT2022upj (\citealt{Chakraborty2025}).
The Eddington ratio for eRO-QPE3 was estimated using only
the QPE bolometric luminosity and the mean black hole mass,
whereas for eRO-QPE4, the mean black hole mass and the combined
disk and QPE bolometric luminosities were used.
}

\end{table}

\subsection{spectral evolution}
\label{sect:spectral_evolution}

Finally, the important element which may help to distinguish between the two scenarios, is the spectral evolution during the QPE event. In GSN 069 the trend is clear: the amplitude of the outburst is moderate (factor $\sim$ 2.5) in the very soft X-rays, 0.2 - 0.3 keV, while rising to a factor above 100 in the 0.8 - 1 keV (see Fig. 2 of \citealt{miniutti2019}). It looks like a strongly variable Comptonization of the seed black body photons, whose spectrum is fitted with the temperature of $kT \sim $ 50 eV. As showed by \citet{arcodia2022} for eRO-QPE1, the outburst starts at lower energies. Spectral complexity of 
'Ansky' \citep{2025NatAs...9..895H,2025hernandez} confirms the same pattern, with the temperature rising during the outburst and decreasing afterwards. The data quality does not allow to distinguish firmly between an expanding medium that emits black body radiation and the evolving Comptonizing medium, although the excess in the model below 0.4 keV might suggest Comptonization (see Fig. 4 of \citealt{2025hernandez}). What is far from being clear is the level of change in the soft X-ray bans related to the bolometric change. In GSN 069 \citep{miniutti2019} and  eRO-QPE2 \citep{lu2023} the change in the bolometric luminosity is relatively small but we do not have reliable estimates of this aspect in other sources.

\section{Basic types of the accretion disk instabilities}

There are two very different families of instabilities expected in accretion disks. The first one is associated with the dynamical evolution of a misaligned disk and may disrupt the disk into two or more almost disconnected parts. Such a situation can arise when newly accreted material is strongly misaligned with the spin of the central supermassive black hole. Frame dragging induces differential Lense--Thirring precession of the inner disk, while internal stresses tend to communicate the warp and maintain a coherent disk structure. Depending on the disk parameters, this competition may result in a smoothly warped disk or, for sufficiently strong differential precession, in disk breaking or tearing. The evolution of misaligned and non-planar accretion disks has been of theoretical interest for a long time \citep{papaloizou1983}. Early numerical studies included \citet{larwoodd1996}. Considerable progress towards exploring disk breaking in realistic black-hole accretion flows was made by \citet{nixon_king2012} and \citet{nixon2012}, while later developments included also multi-ring configurations (e.g. \citealt{stuchlik2015}). Finally, disk tearing has been proposed as an attractive mechanism potentially relevant to the QPE phenomenon \citep{raj_nixon2021}.

The second family consists of thermal and viscous instabilities which do not disrupt the disk geometrically but can produce large-amplitude, and in some cases recurrent, variations of the accretion flow. These instabilities are associated either with the partial ionization zone in the disk or with the dominance of radiation pressure over gas pressure. Ionization instability was first developed in the context of cataclysmic variables \citep{meyer1981,Smak1984}, while thermal and viscous instabilities of radiation-pressure-dominated Shakura--Sunyaev disks \citep{ss73} were recognized soon afterwards \citep{pringle1973,lightman1974}. For some time, these instabilities posed a major difficulty for standard disk models. However, they have been proven to operate in canonical microquasar, GRS 1915+105 \citep{belloni1997, janiuk2000}, and also suggested for reactivation mechanism in quasars \citep{czerny2009}. More recently, radiation-pressure instability has also been discussed in the context of CL AGN and QPEs \citep{sniegowska2020}.

\section{Dynamical instabilities}

Accretion disks do not always have a single, well-defined plane. Non-axisymmetric forces acting in the system can lead to much complex flow pattern. In the vicinity of massive black hole, this could happen if the supply of the matter does not have a single well-defined angular momentum. Even if the accreting material has a well defined angular momentum but the corresponding plane is not co-aligned with the spin of the central black hole, this will lead to the formation of the warped disk. 

Planar disks, like in the standard models of \citet{ss73} evolve under the action of turbulent viscosity, and the nature of this viscosity is related to the magnetorotational instability suggested by \citet{balbus1991}. This instability leads to the development of the saturated turbulence, and the net result is the local dissipation of the energy as well as the transport of both the energy, and more importantly, the angular momentum radially. The angular momentum has just a single component, parallel to the symmetry axis of the system. The system can develop a stationary flow if the matter is continuously provided, and the predictions of such a stationary model, for geometrically thin and optically thick Keplerian disk, emitting locally as a black body  do not depend on the specific form of viscosity since all is set by the energy and angular momentum conservation laws.

In the case of a warped disk, the angular momentum distribution is three-axial, which complicates the picture and leads to new phenomena. Here viscous forces also play a role, and the evolution is governed  by both viscous and dynamical terms (for an excellent introductory material, see \citealt{Nixon_review_2016}).  The warp evolves, and the evolution pattern depends on the relation between the dimensionless viscosity $\alpha$ and the disk height $H$ to the radius $R$ ratio. 

If the viscosity is relatively low, 
the pressure forces dominate and the warp can propagate to large distances as waves. Effective propagation is possible when the disk is close to Keplerian  \citep{wijers1999}, and then the viscosity satisfies: 
\begin{equation}
\alpha  \lesssim \frac{H}{R}
\qquad \Rightarrow \qquad
\mathrm{bending\ wave\ propagation}
\end{equation}

If the viscosity is large then diffusion is at work. In that case both the disk shape and the surface density evolve. The details depend significantly on the viscosity, as well as the boundary conditions.

\begin{equation}
\alpha \gtrsim\ \frac{H}{R}
\qquad \Rightarrow \qquad
\mathrm{diffusive\ warp\ propagation}
\end{equation}

A warped disk does not necessarily become dynamically disrupted. If the internal torques are sufficiently efficient, neighbouring annuli remain coupled and the disk can maintain a smooth, although non-planar, configuration. However, in a strongly misaligned disk around a spinning black hole the differential Lense--Thirring precession increases rapidly towards small radii. If the induced precession becomes too strong for the internal stresses to maintain coherent communication between neighbouring disk annuli, the smooth warped configuration can become unstable.

In this regime the disk can break into distinct annuli or rings which subsequently precess rather independently. This phenomenon is usually referred to as disk breaking or disk tearing. Numerical simulations show that tearing is favoured in geometrically thin disks, for large misalignment angles, and for lower values of the viscosity parameter, although the detailed stability criterion depends on the nonlinear response of the warped disk \citep{nixon2012,rajetal2021}.

The QPE application therefore relies specifically on the tearing regime rather than on the presence of a smooth warp alone. If tearing develops close to the inner edge of the accretion disk, the resulting independently evolving inner disk can display quasi-periodic behaviour and produce strong variations of the accretion flow. This motivated \citet{raj_nixon2021} to propose disk tearing as a possible mechanism responsible for QPEs.

The model is based on the assumption that the angular momentum of the outer disk is misalligned with respect to the spin axis of the central black hole. The evolutionary timescale of the disk itself is the viscous timescale, as in the standard disk. Using the kinematic viscosity $\nu$ prescription we can write
\begin{equation}
    \tau_{visc} = {R^2 \over \nu},
    \label{eq:tvisc}
\end{equation}
where the knematic viscosity is related to the $\alpha$ of \citet{ss73} as $ \nu = \alpha c_s H$, where $c_s$ is the sound speed. This has to be compared with the Lense-Thirring nodal precession timescale which depends on the black hole spin parameter, $a$, at a given radius, and is given by:
\begin{equation}
\tau_{np} = {1 \over 2 a}({R \over R_g})^3 ({GM \over c^3}).
\end{equation}
The viscous and nodal-precession timescales characterize the subsequent variability of a torn disk, whereas the onset of tearing itself is controlled by the nonlinear stability of the warped disk and by the competition between the externally imposed differential precession and the internal torques maintaining radial coherence. 

Differential precession increases the relative positional angles of the newly formed rings. Ring interaction leads to shock formation, sudden local transfer of the angular momentum and subsequent transient inflow of the mater which leads to a flare-like oscillation. The tearing process in the disks has been modeled recently in a number of works \citep{liska2023,liska_corona2023,Asahina2024,speicher2025,fragile2026}.

Numerical solutions of the time evolution depend on the details of the setup. The simulations of \citet{rajetal2021}, for example, show cyclic tearing of the innermost disk, with successive rings breaking off, precessing and subsequently losing angular momentum through interaction with neighbouring rings. When rescaled to the estimated black hole mass of GSN 069, the resulting recurrence timescale is of the same order as the observed QPE period, although no detailed parameter fitting was attempted.

The model seems a natural explanation for a QPE in a TDE source since the original plane of motion of the disrupted star is likely to be inclined with respect to the spin axis of the black hole. However, further modelling is needed to explain why the QPE events in GSN 069 were observed only several years after the TDE. An attractive feature of the tearing interpretation is that the instability is sensitive to the disk thickness. Numerical simulations show that geometrically thinner disks are considerably more prone to tearing, whereas thicker disks may remain coherent. Since the disk thickness generally increases with accretion rate, this provides a possible explanation for the delayed appearance of QPEs in GSN 069: shortly after the TDE the relatively high accretion rate may have maintained a thicker, stable disk, while subsequent decline of the accretion rate allowed the disk to become sufficiently thin for tearing to develop. This possibility has been noted in the context of GSN 069, although a time-dependent model reproducing quantitatively the observed transition has not yet been presented. Detailed studies of the disk evolution even prior to QPE could help to evaluate the full potential of the model.

\section{Thermal/viscous instabilities}

Shortly after the formulation of a very convenient and simple theory of stationary accretion disks by \citet{ss73} it was discovered that those models correspond to unstable solutions. Thermal instability was analyzed by \citet{pringle1973}, and viscous instability by \citet{lightman1974}, and it was shown that the two instabilities are actually closely related \citep{ss76}. The instability is present in the disk region which is dominated by the radiation pressure, i.e. it is expected in high luminosity sources. As the contribution of the radiation pressure at a given Eddington ratio depends on the central black hole mass, the instability should be more important for more massive black holes than for lower mass sources like those in X-ray binaries. On the other hand, at very high Eddington ratios the instability can be suppressed by the rapid inflow into the black hole which serves as a cooling mechanism, so such disk models (slim disks; \citealt{abramowicz1988}) are stable.

 When the Eddington ratio approaches unity, the role of the radiative pressure rises, the radial extend of the radiation-pressure dominated zone increases, the role of the radial advection term is rising. The vertical structure (in particular, the disk thickness to the current radius ratio, $H/R$) changes rapidly as 
$\lambda$ approaches unity. In addition, the likelihood of strong winds is also rising.  The second remarkable limit is at $\lambda $ at the level of a few per cent, when the inner optically thin flow may eventually develop. The dynamical instabilities leading to tearing the disk are also sensitive to the disk thickness. Thus, estimates of the $\lambda$ are very important for constraining the QPE mechanism.

The local thermal equilibria frequently depicted as S-curves 
\citep{1998kato} 
at fixed radii, where the advection cooling term gives an upper branch. The curve is then composed of three parts: a lower Shakura–Sunyaev thin disk branch, an intermediate (unstable) radiation-pressure region, and an upper, advective, slim-disk branch that is thermally and viscously stable. 
\begin{figure*}
    \centering
    \includegraphics[width=0.85\columnwidth]{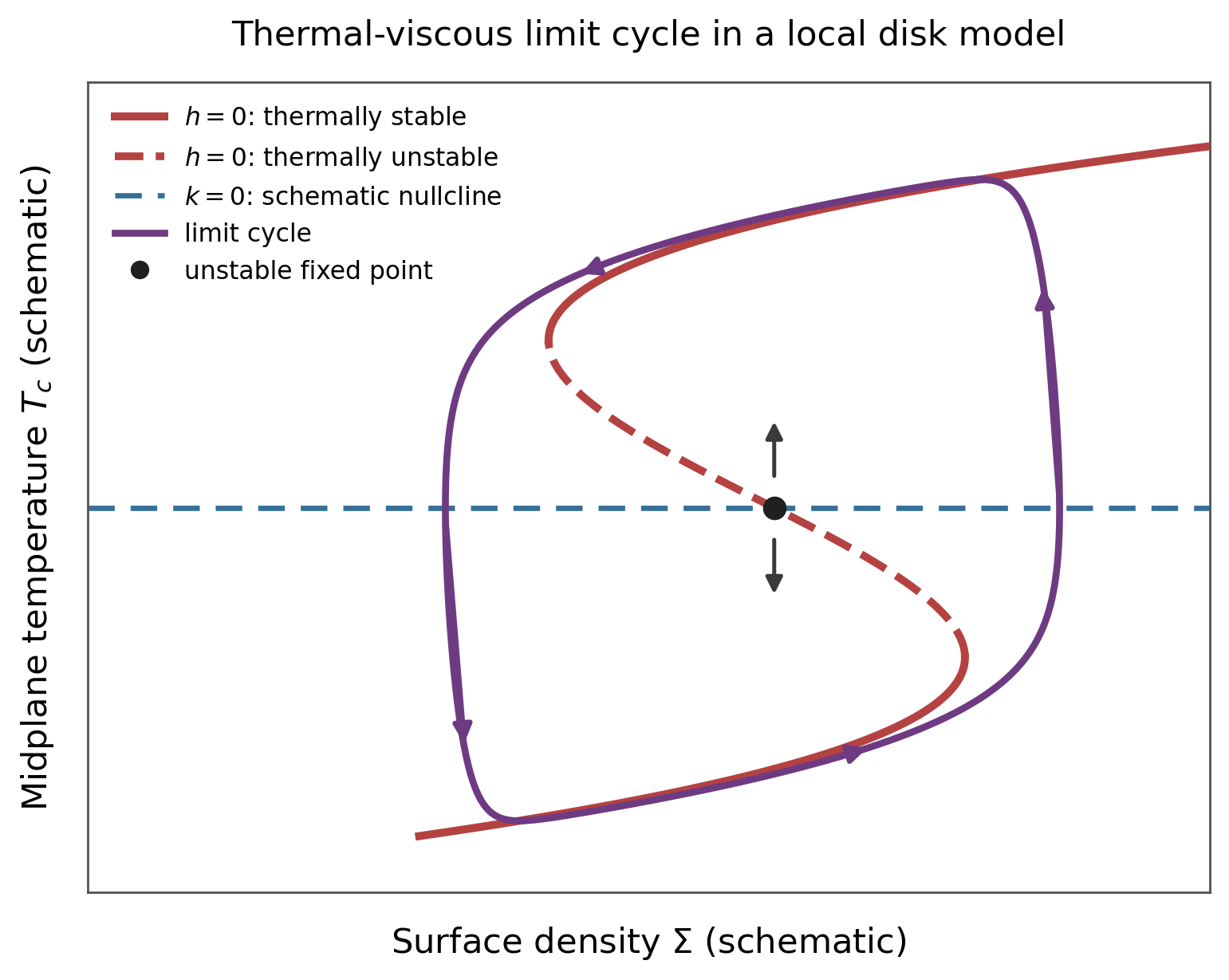}
    \caption{Schematic phase portrait of a local thermal–viscous instability in an accretion disk. Solid and dashed red curves indicate thermally stable and unstable branches of the thermal equilibrium curve (\(h=0\)), respectively. The blue dashed line represents the surface density nullcline (\(k=0\)), whose intersection with the equilibrium curve defines an unstable fixed point. The purple trajectory illustrates the resulting limit cycle, characterized by rapid thermal transitions at nearly constant surface density and slower viscous evolution along the stable branches. The axes ranges are schematic and do not represent a particular physical scaling.
    } 
    \label{fig:scurve}
\end{figure*}

In Figure \ref{fig:scurve} we show schematic view of vertically integrated accretion disk local solutions, i.e. the phase space of surface density - mid-plane temperature. The stability of thermal equilibria, represented by this curve, may be studied by imposing small perturbations of density and temperature over time. In the equilibrium the following balance is satisfied  when $h=0$:

\begin{equation}
   \Sigma T  \frac{\partial S}{\partial t} = Q^{+} -  Q^{-} = h(T,\Sigma)
\end{equation}

where $Q^{+}$ and $Q^{-}$ denote vertically integrated heating and cooling terms,  with $Q^{-}$ containing radiative and advective cooling, and $S$ being the specific entropy.
At the same time the surface density evolution is governed by diffusion equation, expressed in following way:

\begin{equation}
{\partial \Sigma \over \partial t} = C_{\rm thv} k(T,\Sigma)
\end{equation}
where $C_{\rm thv} = (c_{\rm s}/v_{\phi})^{2}$ is the ratio of local thermal and viscous timescales and $k$ represents the net rate of surface density evolution in the local model.
The solution of disk structure, for a given accretion rate, is a fixed point, that lies on the intersection of $h$ and $k$ lines.
so that $Q^{+}=Q^{-}$ and $T=T_{0}$ are satisfied. On the stable branch, the fixed point is attracting, as any small perturbation in temperature is damped: if temperature rises slightly, the system moves along the $h=0$ line, on the thermal timescale, to find a new value of $\Sigma$, which evolves on viscous timescale. This new $\Sigma$ will define new $k=0$ intersection, but then the small change in accretion rate, corresponding to the new temperature value, will bring the system back to the old value of density (it will drop, if temperature rises). System will therefore come back to the attracting fixed point solution, and perturbation is damped.
On the other hand, the unstable branch contains repelling fixed points. Here any small perturbation in temperature will lead to a runaway heating, with surface density decreasing on viscous timescale, and accretion rate increasing. The instability may lead to break up of the disk and its destruction, if not controlled by existence of the upper stable branch of equilibria. The one which stabilizes the solutions for the radiation-pressure dominated disks at high $\dot M$, is the advective branch. As a result, in the range of accretion rates corresponding to negative (unstable) slope of the S-curve, the disk will oscillate between cold and hot states, and limit cycle will be operating.

This sequence of local solutions on the temperature-surface density plane is closely analogous in mathematical structure to the other well known S-curve representing the disk instability model (DIM) for dwarf novae \cite[eg.][]{2000Hameury, 2002lasota}.
It should be noted that the physical origin of the two instabilities is different, and in DIM it is the partial ionisation of Hydrogen that affects the thermal balance due to the opacity inversion. In consequence, the regions of temperatures around $\log T \sim 3.5$ are unstable, corresponding to the ${\partial \Sigma \over \partial T} < 0$ branch.
The region of partial Hydrogen ionization exists in most accretion disks, but its location depends on the accretion rate, and the unstable strip shifts outwards for higher $\dot M$ (i.e. hotter disks).  
The regions unstable due to radiation pressure are typically extended from the innermost stable circular orbit (ISCO) up to several hundreds of $r_{\rm g}$, also depending on the accretion rate. However, for cool disks (small  $\dot M$) this region shrinks and all disk is gas-pressure dominated.

We notice that for WD, BH and NS, both regions of instabilities due to radiation pressure and Hydrogen ionization can exist in the same disk, if only the accretion rate is high and the disk size,
that is limited by Roche lobe size, is also large. For WD, the radiation pressure zone is not
extending beyond the WD radius, so in practice we will not observe this kind of instability.

Below, we reproduce here the plots from 
\cite{janiuk2011}, albeit for a different choice of parameters, more adequate for the intermediate-mass black holes.
The plot is result of numerical calculations of sequence of steady-state models that provide initial and boundary conditions for the code GLADIS - global accretion disk instability simulation \cite{2020janiuk}.

\begin{figure*}
    \centering
    \includegraphics[width=0.48\columnwidth]{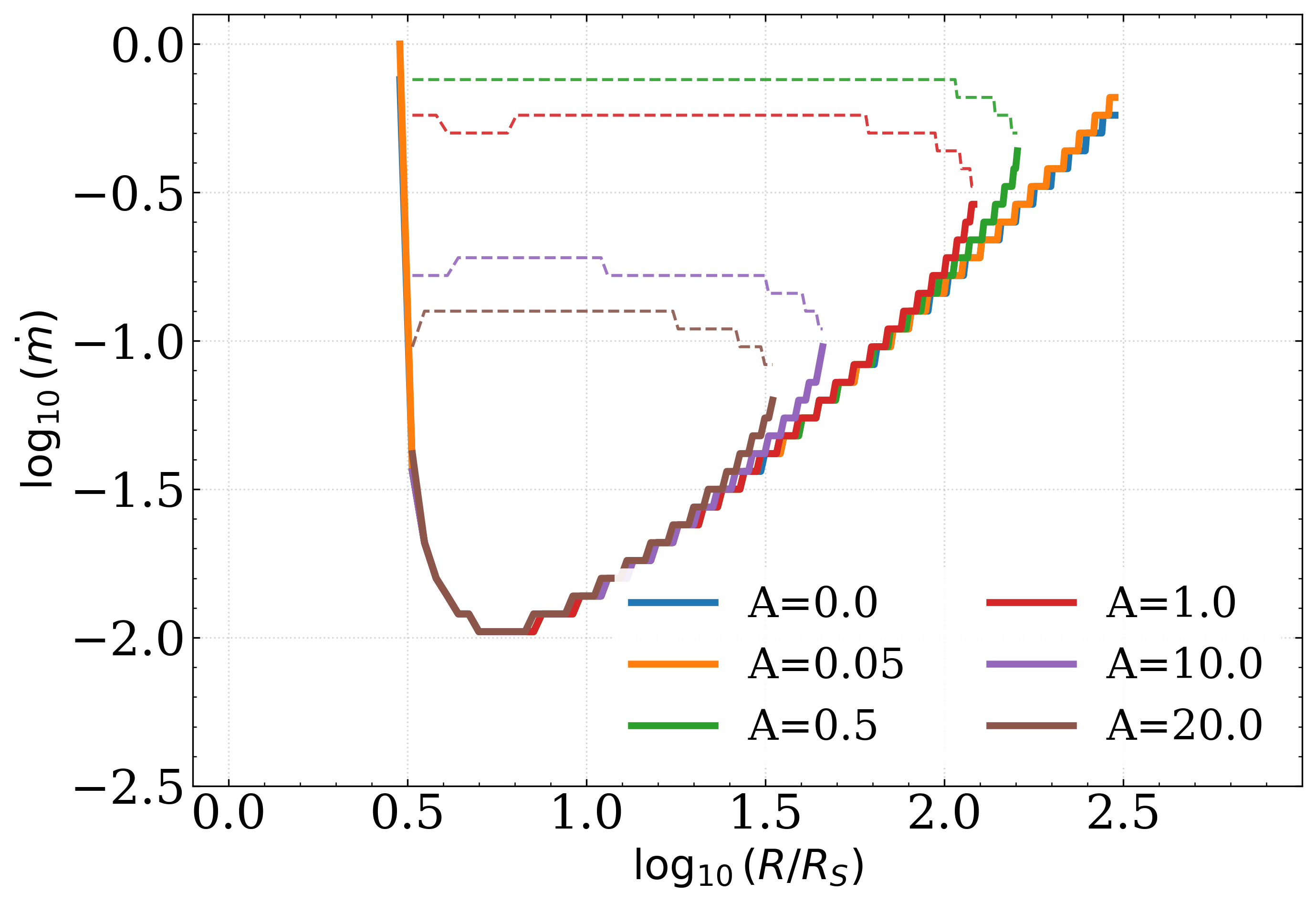}
      \includegraphics[width=0.48\columnwidth]{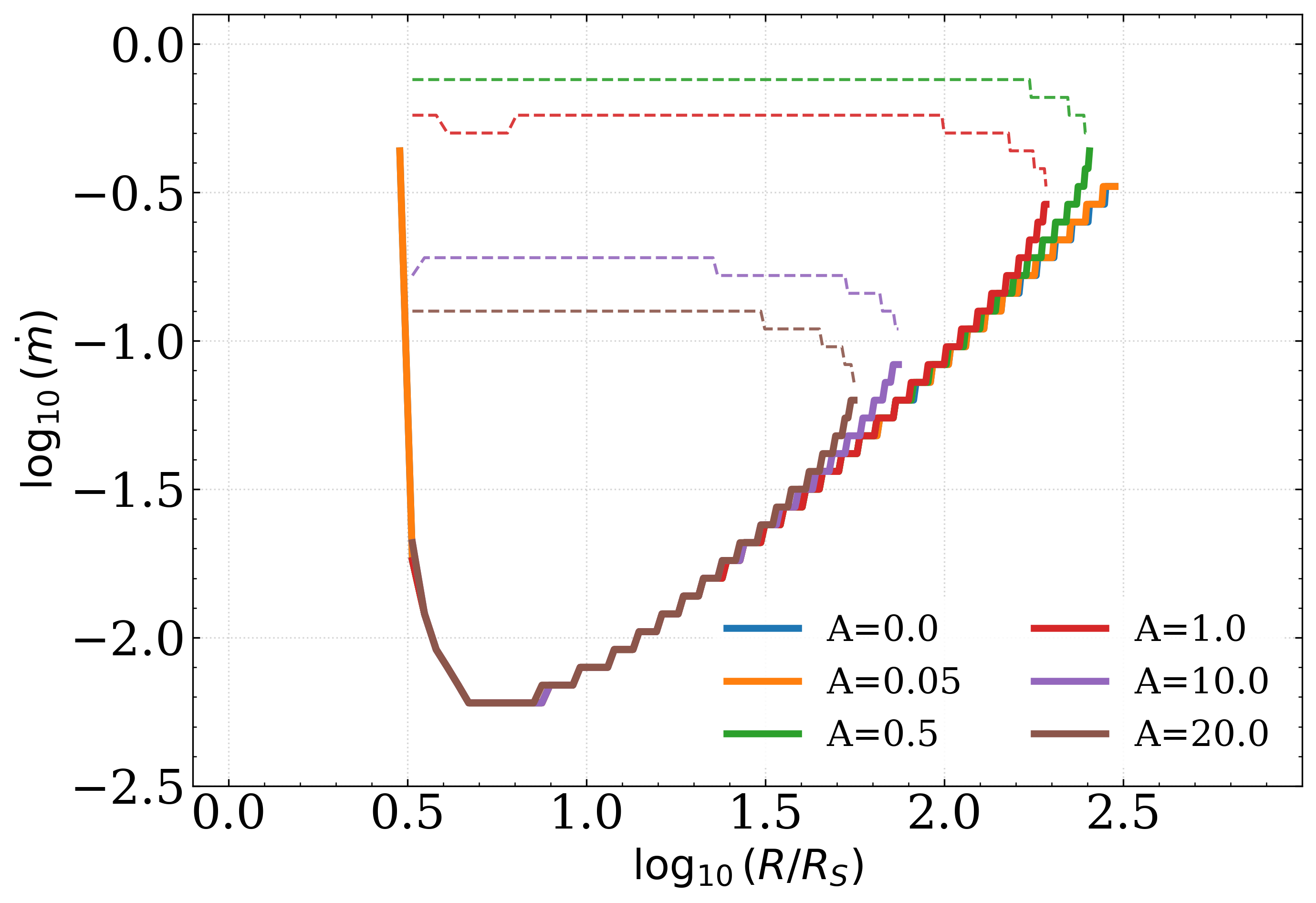}
    \caption{ The spread of instability zone calculated using the Scurves for two cases of intermediate mass black holes $M_{BH} = 10^4 M_{\odot}$ (left) and $M_{BH} = 10^6 M_{\odot}$ (right), shown for various wind outflow parameters. Solid line shows the instability starting point and dashed line shows the ceasing point for a particular radius in the disk. Computations of GLADIS are performed between R$_{in} = 3R_{schw}$ and R$_{out} = 300 R_{schw}$}. 
    \label{fig:instability zone}
\end{figure*}

It was found (e.g., \cite{2015A&A...574A..92J}) that the wind outflow from the accretion disk serves as an important stabilizing mechanism, affecting the overall spread of instability in the disk. In the current version of GLADIS code, the wind parameterisation has been included as a quadratic function of the local accretion rate (in dynamical evolution, it is dependent upon the time and radius), as:
\begin{equation}
f_{out} = 1 - {{1} \over {(1+A {\dot m}^{2})}}.
\label{eq:fout}
\end{equation}
where "A" is the outflow parameter and "$f_{out}$" is the energy lost from the disk due to the outflow, factored in the radiative cooling of the disk, as in \citet{Nayakshin_2000}.
In Figure \ref{fig:instability zone}, we show the effect of this wind outflow parameter on the spread of instability zone in the disk for two different intermediate mass black holes. As expected, stronger wind reduces the extent of unstable zone in the accretion disk.

\section{Radiation pressure instability - advantages and drawbacks of the model}
\label{sect:model}

The advantage of the radiation pressure instability mechanism is that it has a reasonable predictive power. The 1-D codes well describe the time evolution of an vertically averaged disk, with or without a hot corona, and with or without wind/jet outflow \citep{janiuk2000,janiuk2002,janiuk2011,grzedzielski2017}. The conditions triggering the instability involve an assumption about the scaling of the viscous torque with the conditions inside the disk, i.e. gas pressure, radiation pressure and magnetic pressure. This scaling affects the stability (scaling of the torque with gas pressure only leads to a stable solution, \citealt{sakimoto1981}. 
 
Most Galactic black hole binaries maintain stable high/soft disks where the $\alpha$ disk model and scaling with total pressure 
predicts the instability. However, the heartbeat variability is not limited to GRS 1915+105
and IGR J17091-3624. It is also observed in other sources, e.g. 4U 1630-47 \citep{Fan2025}. Moreover, recurrence analyses additionally
identified nonlinear variability in XTE J1550-564 and GX 339-4 \citep{2016sukova}, although this is not
unique evidence for radiation-pressure limit cycles. The relevant question is why instability operates only in
some systems. It was shown already that magnetic support, alternative stress prescriptions, and winds can alter or suppress it \citep{janiuk2011, 2015A&A...574A..92J}.

\subsection{Basic parameters setting the evolutionary timescale of the accretion disk}

In the case of AGN, also arguments based on the episodic (intermittent) variability point towards the presence of the radiation pressure instability \citep[e.g][]{czerny2009,schawinski2015}. However, those timescales are of order of thousands of years, unlike QPE events. 
 Without additional parametrization in the instability mechanism, the recurrence timescale is primarily led by the viscous evolution of the unstable region (negative slope of the S-curve).
The viscous timescale \[ t_{\rm visc} \sim \alpha^{-1} \left(\frac{H}{R}\right)^{-2} \Omega_{\rm K}^{-1} \] 
{\bf cf Eq. \ref{eq:tvisc}} is the timescale of an inflow of the material in the Keplerian disk, with $\alpha$ as the viscosity parameter, $H/R$ as the hight to radius ratio, and $\Omega_{\rm K}$ the Keplerian angular velocity. Thus, the parameters that can shorten this timescale are  a larger viscosity $\alpha$, a larger $H/R$, or smaller characteristic radius of the unstable region. 
Increasing the viscosity parameter results in a shortening of the flare period without significantly changing its amplitude (changing $\alpha$ shifts S-curve in horizontal direction), as shown, for example, by \cite{janiuk2000} for 10$M_{\odot}$ black hole;
see also \cite{sniegowska2020} and their Figures 3 and 4 for a $10^7 M_{\odot}$ accretor.
Increasing the disk thickness \citep{pringle1981} would shrink the timescale as well, since the material is faster transported inward in a geometrically thicker disk.
If the instability region can be shrunken down to very narrow radial range the the timescales can be much shorter as well.
The idea that a smaller part of an accretion disk can evolve faster than the disk as a whole, resulting in shorter outburst timescales, was discussed in the context of GRS 1915+105 by \cite{belloni1997}. In their scenario, the inner disk region can rapidly evacuate and then it is refilled by matter flowing inwards from larger radii. 
An alternative scenario where the inner hot flow is surrounded by an outer narrow instability zone (expected for the Eddington ratio marginally allowing for the radiation pressure instability) was suggested by \citet{sniegowska2020} for CL AGN. More detailed 1-D numerical computations by \citet{sniegowska2023} showed, however, that the outburst started in a narrow zone still propagates outwards, and the time scales remain long. Minor numerical code modifications done for that paper did not affect much this conclusion, and the only parameter which did shorten the outburst dramatically was the assumption of a very small outer disk radius.

In the context of QPE, the presence of such a small disk radius may be justified. In these sources we have a TDE, so either the disk just forms from the disrupted star, or, in case of TDE in previously active nucleus, the TDE cloud destroy the old disk \citep{2024ApJ...973..101L}.
Therefore, the outer radius of order of tens of Schwarzschild radii can be justified. Such small outer radius both reduces the timescales to a fraction of a year, and also leads occasionally to irregular outbursts as the propagating waves are scattered at the disk outer edge (see Fig. 12 and 13 of \citealt{sniegowska2023}). 

\subsection{Instabilities with modified viscosity}

It can be shown that modified viscosity laws, such as the one used by \cite{Nayakshin_2000} and \cite{2007A&A...466..793J} and scaling the viscosity locally with the gas-to-radiation pressure ratio; i.e.:
\begin{equation}
\alpha = \alpha_{0} \frac{1+\xi/\xi_{0}}{1+(\xi/\xi_{0})^{2}}
\end{equation}
where $\xi=P_{rad}/P_{gas}$ will lead to somewhat shorter variability timescales (the regular outbursts appear about 10 times more frequently, for typical parameter, $\xi_{0}=8-10$).
Moreover, the propagating viscosity fluctuations \citep{2012A&A...540A.114J} will stabilize the disk partially or completely, and leave only stochastic type of variability, instead of large amplitude prolonged flares. Also, scaling the stress tensor with geometric mean of gas and total pressure, instead of the total pressure, reduces instability range \citep{janiuk2011}.

Models based on a more general approach to the viscosity prescription for the tensor $\tau_{r \phi}$
\begin{equation}
\tau_{r\phi} = \alpha P_{gas}^{\mu}P_{tot}^{1 - \mu}
\end{equation}
were also considered \citep[e.g][]{szuszkiewicz1990,merloni2006}.
With this modified prescription \citep[see][]{grzedzielski2017}, the viscous stress responds to pressure perturbations in a phenomenologically different way than in the standard $\alpha$-disk model.

\citet{grzedzielski2017} showed that most reasonable models are obtained for the dimensionless parameter $\mu \sim 0.5$. With increase of $\mu$, the unstable portion of the S-curve shrink which affects also the amplitude of flares and $\mu$ is value between 0 (gas pressure dominates) and 1 (radiation pressure dominates). 
These viscosity laws are phenomenological stress prescriptions, not predictions of
MRI turbulence; notice that $p_{tot}$ here does not include magnetic pressure -- is is a sum of gas and radiation pressures. The timescales are shorter than expected for $\mu \sim 0$ but still of order of hundreds to thousands of years. Models predict rather regular outbursts, with shorter duration and longer separation between the outbursts, particularly for lower Eddington ratio \citep{janiuk2000}. Luminosity at the peak is also quite high since the operating limit cycle is completed at the hot branch by advection. Too high luminosity and too long timescales seemed to prevent the application of the model to QPE.

\subsection{Instabilities with magnetic field}

 Magnetically supported disks are in general geometrically thicker due to the additional vertical support provided by magnetic pressure. This may shorten the viscous timescale and also reduce the radial extent of the radiation pressure dominated unstable region \citep{begelman2007, dexter2019}.
 Hence, MRI-driven stresses and vertical support does not need to follow the $\alpha$ viscosity prescription, while the RMHD calculations find the existence of thermal runaway \citep{2013jiang,2016Mishra}, where dissipation and radiation advection may occur out of the mid-plane. Other simulations find stable
magnetically supported configurations \citep{2016sadowski, 2025zhang, 2025ApJ...988...43J, 2025blaes}. \cite{Kaur2023} showed that magnetic support can shrink the unstable region and inferred QPE
periods, but these authors did not evolve global model and did not obtain recurrent light curves. 
\citet{Alush2025} employed the model of \citet{Kaur2023} in 1-D time-dependent computations but they concentrated mostly on self-similar solutions which do not catch all the aspects of time-dependent behaviour.
The GLADIS simulations, where most recently \cite{sniegowska2023} added the extra magnetic transport as a cooling component to the thermal energy conservation, follow long-term evolution. Even though they adopt a phenomenological magnetic field prescriptions which would require further calibration against recent MHD results, these simulations were able to present observables in the form of long-term lightcurves. 

Making outburst timescales to match QPE which are very short seems possible but the corresponding modification has not been done to any time-dependent code. But the timescale analysis of the significantly modified disk model suggests it is possible. By employing the large-scale magnetic field and the nonzero torque condition, \citet{Pan22,Pan23} showed qualitative agreement with not only the timescales and energetics of most QPE but also the the phase-resolved X-ray spectrum, though amplitude changes in outbursts (i.e. in  eRO-QPE1 \cite{arcodia2021}) remained challenging to explain by disk instability models. Recently, \citet{Pan25} revealed that QPE can be divided into two categories by the critical thresholds of mass accretion rate and magnetic field parameter. When the mass accretion rate or magnetic field parameter is smaller than the critical thresholds, QPE (e.g., GSN 069) will stay at the stable regimes and manifest as regular eruption patterns. Otherwise, QPE (e.g., eRO-QPE1) will fall into the unstable regimes and their periods will be highly stochastic. This framework can naturally explain the two different eruption patterns in QPE while keeping the peak temperature nearly constant.

The relation between the period and the magnetic-field coefficient is non-linear, as changes in the magnetic-field strength eventually lead to a different outburst morphology (see Fig. 15 of \citealt{sniegowska2023}) using effective modification of the disk zone due to the action of the magnetic field (Equation 23 in \citealt{2015begelman} ).
In \cite{sniegowska2023} we showed that including a magnetic field into our simulations can shorten the outburst timescales by a factor of 25 in comparison to the model without a magnetic field, before it stabilizes the inner disk. However, the addition of magnetic field affects the amplitude of the flare by factor of 20-25.
As \cite{Pan23} showed, it is possible to tune parameters of the model to reproduce the shape and timescales of some of QPEs, however high $\alpha(\simeq$ 0.1) is needed to do so. Proper time-dependent computations are still to be done to proof it.

\subsection{Puzzle of the period changes}

A new important phenomenon worth noting in QPE is the continuous increase  of period in Ansky, of which the period varies from about 4.5 days in 2024 \citep{2025NatAs...9..895H,2025hernandez} to roughly 14 days in 2026 \citep{2026Chakraborty}. While this smoothly increasing period remains challenge for the star-disk collision model, it can be easily explained by the disk instability model with increasing mass accretion rate. Furthermore, it is interesting that the outburst amplitude in 2024-2025 has a positive relationship with the period. This result is in conflict with the star-disk collision model since the larger period implies a larger orbit radius and smaller rotational velocity, leading to the smaller outburst amplitude. However, this positive relationship is inverse in this year's observations. Faint outbursts are observed while the periods keep increasing, indicating the complexity of this source.

Summarizing, the timescales predicted by the radiation pressure instability seem for the moment too long to explain well all the QPE unless some specific assumptions about the viscosity mechanisms are made. The TDE seems to be required, while the radiation pressure instability may operate on a very short timescale, if a small outer disk radius is assumed. This is technically needed to prevent the outburst propagation outwards. However, the TDE mechanism gives no constraints for reviving the QPE activity a number of times, which may be a problem for partial TDE and returning debris.

\section{Limitations of global RPI models and possible role of a warm corona}
\label{sect:limitations}

\begin{figure*}
    \centering
    \includegraphics[width=0.32\columnwidth]{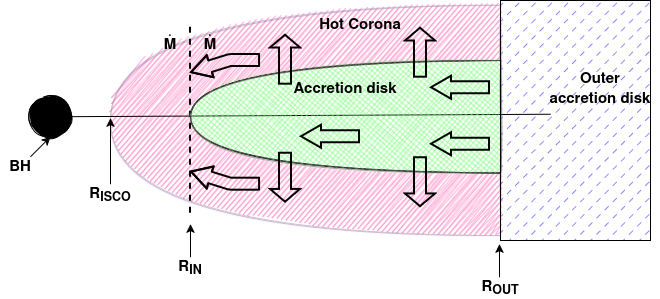}
    \includegraphics[width=0.6\columnwidth]{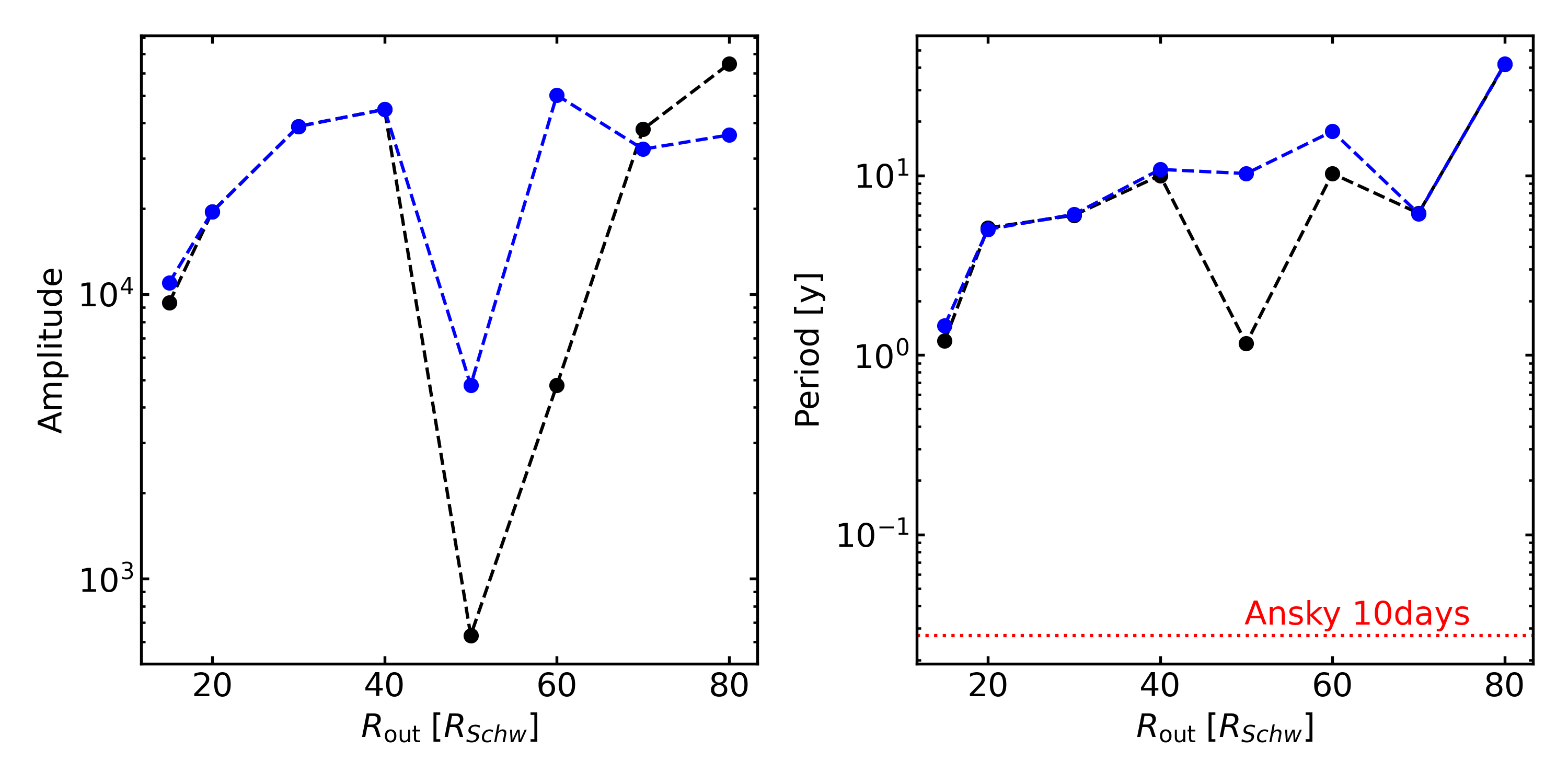} 
    \caption{Left panel: Schematic view of the innermost part of the flow in the recent version included in GLADIS code.
    Accretion disk (green), and hot corona above the disk (red), outer, stable part of accretion disk we marka purple dashed box. The horizontal black line represents the equatorial plane. The vertical dashed line represents the boundary condition between the ADAF and outer flow. Computations of GLADIS are performed between R$_{in}$ and R$_{out}$ marked in the drawing. Dependence on the amplitude of the outburst (middle panel) and the duration of the limit cycle (right panel) for different short outer radii of the accretion disk for the black hole mass 10$^6M\odot$ and $\dot m = 0.14$. As black points we mark model without hot corona, as blue points we mark model with hot corona included. The symbols mark the model results. The dashed lines connecting the points are included only to highlight the trend between the computed models. As dotted red line, we mark one of observed periods of Ansky \citep{2025hernandez}.  }. 
    \label{fig:our_model_combo}
\end{figure*}

Global computations with the code GLADIS\footnote{\url{https://github.com/agnieszkajaniuk/GLADIS}} \citep{2020janiuk} show the limitations of the current RPI model but also indicate future prospects. Studies models incorporated an effect of the hot corona horizontal layer as shown in Figure~\ref{fig:our_model_combo}, left panel (see details in \citealt{sniegowska2023}.) Such a geometry is consistent with the recent arguments based on polarization \citep{Ingram2023,Gianolli2023} In  \cite{sniegowska2023} this corona was formed due to magnetic buoyancy, but the accreting hot corona did not shorten the outburst timescales considerably. The only effective way to shorten the recurrence time was to reduce the outer disk radius. The problem is illustrated in Figure~\ref{fig:our_model_combo}. Here the disk is not dominated by the magnetic field as proposed by \citet{Kaur2023} which could shrink the timescales to the required values. However, the interesting aspect of the very small outer radius is the appearance of the irregular outbursts \citep{sniegowska2023} which are caused by the reflection of the cooling/heating wave from the outer disk edge. Full 1-D computations for magnetically-dominated models would show if these models lead only to smooth and regular outbursts.

\begin{figure*}
    \centering
    \includegraphics[width=0.49\columnwidth]{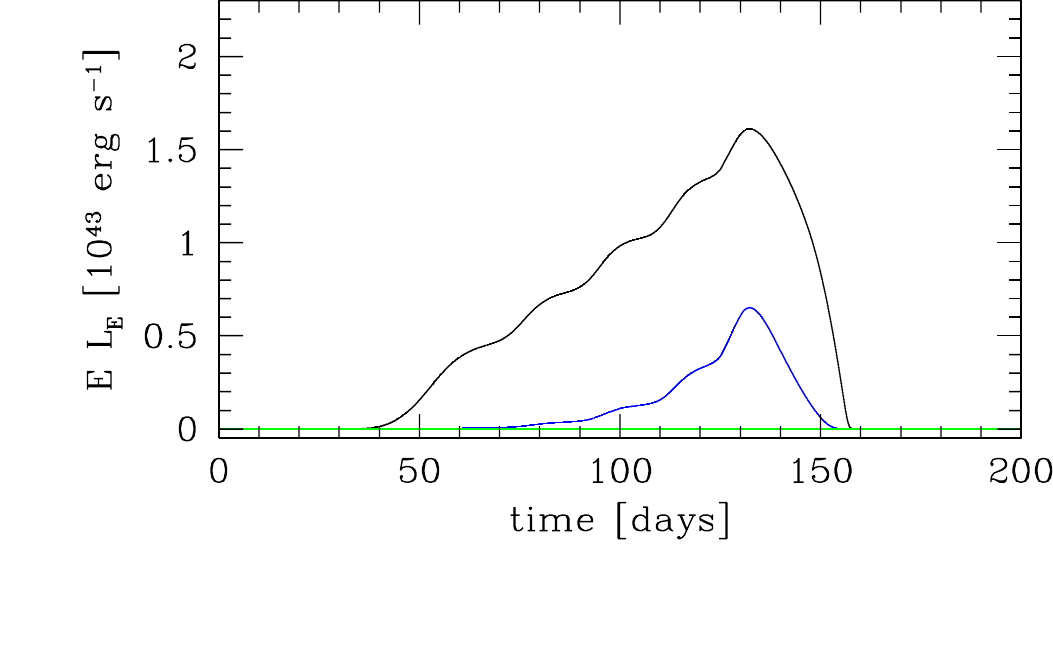}
     \includegraphics[width=0.49\columnwidth]{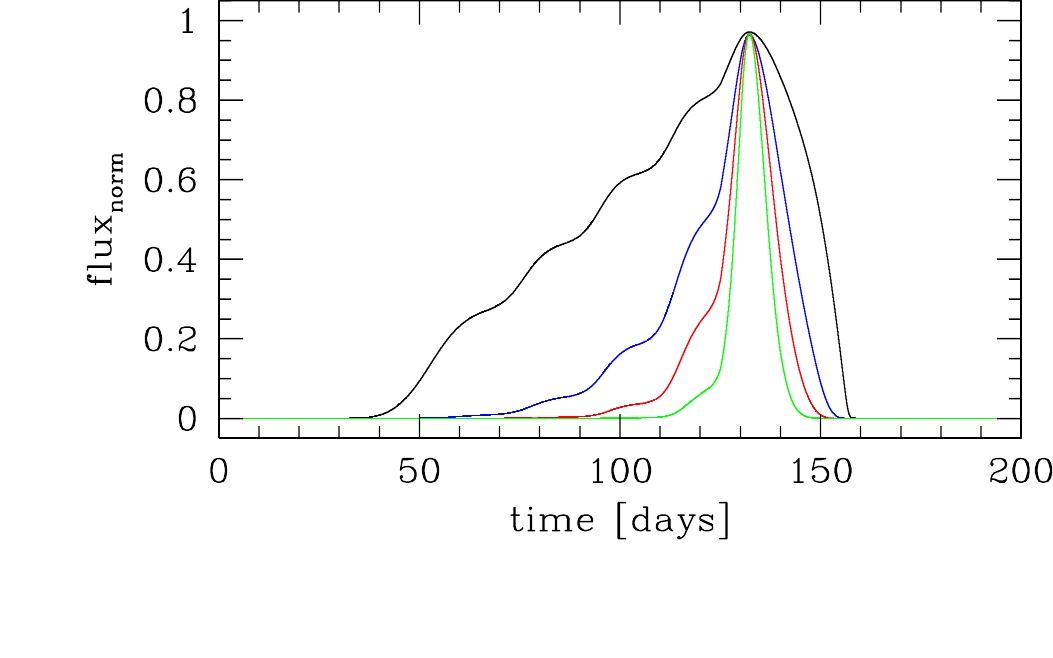}   
    \caption{ The time profile of the disk flux at 0.05 keV (black line), 0.2 keV (blue line), 0.4 keV (red line) and 0.8 keV (green line) for the outbursts assuming the black hole $10^6 M_{\odot}$, dimensionless accretion rate $\dot m = 0.14$, the inner radius at $3 R_{Schw}$, the outer radius at $15 R_{Schw}$,  with flux in physical units (left panel) and flux normalized to the peak of each lightcurve (right panel). This are illustrative plots which show the general energy-dependent trend observed in GSN 069 (see \ref{sect:spectral_evolution}), although the timescales are longer than observed.} 
    \label{fig:time_energy_profile_new}
\end{figure*}

GLADIS simulations, although they do not offer satisfactory representation of the data, still show some interesting trend. We selected one of the outbursts representing the small disk, with inner radius at ISCO and the outer radius at 15 $R_{Schw}$. Knowing the temperature profile during the outburst and assuming  the black body emission of the disk we could calculate the outburst profile at specific energies. 
The Figure ~\ref{fig:time_energy_profile_new} shows the outburst profile at a few representative energies. We used there $E F_E$ convention which is more suitable for broad instrumental channels.  We see that the outbursts in the disk instability model starts first at lower energies, as requested by the data. The outburst is also generally narrower at higher energies. This pattern is interesting despite the problems with amplitude (the outburst amplitude is too large, particularly at low energies) and timescales. 

Apart from introducing the entire disk to be magnetically dominated we could in the future consider the three-zone model with the magnetically-dominated dissipative warm corona. Such a medium is called for by the spectral analysis as it can explain the soft X-ray excess seen in most AGN \citep[see e.g.][]{walter1993}. This medium is now broadly named 'warm corona', its temperature is of order of a fraction of keV, and the optical depth of order of 10 - 20 \citep[e.g.][]{czerny1994,rozanska2015,petrucci2018,petrucci2020,palit2024,ballantyne2024}. The formation mechanism for this medium is not well known but it is likely related to the suppressed radiation pressure instability' (see the discussion in \citealt{czerny2003}. The medium is shown to be thermally unstable by \citet{gronkiewicz2023}. The medium carries more matter, so its effect on the disk evolution might be much more important than that of the hot corona discussed above. In most recent GRMHD simulations by \citet{lancova2026} for a $10 M_{\odot}$ in a binary system  the formation of the medium resembling the warm corona is actually visible. However, such simulations are not suitable for the global evolution modeling as they are initialized from the torus, with a limited amount of mass and specific boundary conditions. So incorporating this option into GLADIS would be clearly important.

\section{Observational tests of QPE models}

 Further progress in understanding the physical origin of QPEs requires observational tests that can distinguish between accretion disk instabilities and geometrical or orbital scenarios. The basic timescales of the outburst duration and separation can be met now in all basis scenarios, including disk dynamical and thermal instabilities as well as the TDE events followed by the subsequent passages of the stellar remnant through the stellar debris forming a disk. We do not follow the repeating events long enough to rule out any of the proposed mechanisms simply at the basis of simple dynamical requirements like outburst regularity, including systematic changed in the outburst periods. Unexpected trends can be usually explained as a more complex interaction than in the simplistic idea. Small number of sources also poses a problem in any attempt of statistical approach. 

At least two sources seem to have imprints of the luminosity state (and possibly also accretion rate) on their behaviour. GSN 069 showed the disappearance of the QPE after dimming period and subsequent recovery after an increase in the luminosity \citet{miniutti2019}, it was modeled and stressed by \citet{Pan25} as a strong argument for the disk thermal instability. The detailed study of the period in eRO-QPE2 with its numerous detected eruptions has been satisfactorily explained by disk instability models with some random small period changes, but the model of disk crossing with apsidal precession can be also accommodated \citet{Arcodia2026}.

The disk tearing instability also offers a possible connection between QPE activity and the accretion state. In this scenario, differential Lense–Thirring precession can cause a misaligned disk to break into independently precessing rings, leading to recurrent episodes of enhanced accretion \citep{raj_nixon2021}. Since the development of this instability depends on the disk thickness and its ability to communicate warps, changes in the accretion rate may affect the occurrence and recurrence times of the eruptions. However, unlike in some thermal instability models, no robust quantitative relation between the QPE recurrence period and the accretion rate has yet been established for disk tearing. Consequently, the observed connection between QPE activity and the luminosity state does not by itself distinguish between these mechanisms.

Further tests could use the information, whether the changes in the source luminosity correlates with the QPE period, or with the amplitude and fractional duration of the active phase. More observational material and better model predictions in this context should help. The next promising, but even more complex aspect is the spectral evolution.

Time-resolved X-ray spectroscopy covering complete eruption cycles is particularly important. The evolution of the temperature, luminosity, and effective emitting area may help to distinguish between intrinsic changes in the disk structure and transient heating associated with interactions between an orbiting body and the disk. Hysteresis in hardness–intensity or luminosity–temperature diagrams could provide an additional constraint. Indeed, hysteresis in the luminosity–temperature plane has already been observed in several QPE sources, including GSN 069, eRO-QPE2, and AT2019qiz \citep{arcodia2022,miniutti2023,arcodia2024,nicholl2024}. The temperature typically reaches its maximum before the luminosity peak, indicating an evolution of the effective emitting area during the eruption. Spectral state transitions resembling those observed in X-ray binaries have also recently been reported in the non-QPE tidal disruption event TDE 2025aarm \citep{2026arXiv260709850B}. However, these transitions occurred during the long-term evolution of the source, rather than during individual eruptions. Detailed theoretical predictions for the shape and direction of the hysteresis loops in different QPE models are still needed to establish their diagnostic power.

Another promising diagnostic is the variability of absorption features and reprocessed emission. High-resolution X-ray spectroscopy has already revealed a highly ionized outflow in GSN 069, with velocities of about 2000–3000 km s$^{-1}$, and indications that its ionization state is related to the presence of QPE activity \citep{Kosec2025}. More recently, phase-dependent absorption and emission features have been detected in Ansky, revealing a much faster outflow with velocities of about $0.2c$, apparently associated with individual eruptions \citep{Chakraborty2026}. These findings demonstrate that outflows can be an important part of the QPE phenomenon. Further phase-resolved spectroscopy could determine whether changes in absorption are caused primarily by variable ionization, changes in the mass outflow rate, or geometrical effects. In particular, models involving precession of an optically thick disk and wind predict phase-dependent changes in the observed absorption and spectral properties \citep[e.g.][]{Middleton2025}. Simultaneous observations at X-ray, UV, and optical wavelengths could additionally constrain the geometry and timescales of reprocessing, although detecting the expected correlated variability remains challenging. Such measurements, combined with quantitative predictions from different models, may help distinguish between geometrical scenarios, disk–orbit interactions, and intrinsic accretion disk instabilities.

Finally, long-term monitoring is essential for studying the evolution of the recurrence period and the energy released during individual eruptions. In disk instability models, both quantities may depend on the accretion rate and the physical state of the disk, potentially leading to correlated secular variations. Orbital models, on the other hand, introduce an underlying dynamical clock, although orbital evolution and precession may also produce period changes. Measurements of the period drift, its regularity, and its possible correlation with the eruption luminosity and energy output can therefore provide important constraints on the competing scenarios. None of these diagnostics is conclusive on its own, but their combination, together with improved theoretical predictions, should help to identify the physical mechanism responsible for QPEs.

\section{Summary}

Quasi-periodic eruptions remain challenging to explain, despite the rapidly growing observational evidence and the development of several competing theoretical models. The proposed mechanisms can be broadly divided into intrinsic accretion disk instabilities, including thermal and dynamical instabilities, and geometrical or orbital scenarios involving interactions between orbiting bodies and accretion flows. Tidal disruption events (TDEs) may play an important role in providing the accreting material, without necessarily determining the physical mechanism responsible for the eruptions themselves. It is therefore possible that more than one physical process is involved, and different QPE sources need not share exactly the same origin.

Radiation-pressure instability remains a plausible, but not established, explanation. Its development depends on the magnetic support, turbulent stresses, and energy losses in the accretion flow. Classical thermal-viscous instability models tend to predict recurrence timescales longer than those observed in QPEs. Restricting the unstable region to a narrow annulus can reduce the timescale, but may also produce more sinusoidal variability rather than the sharply peaked eruptions typically observed. Recent models incorporating magnetic fields, modified disk geometry, and additional emitting components offer promising improvements, although some assumptions remain insufficiently constrained by accretion theory. In particular, detailed predictions for the time-dependent spectra are still needed.

Dynamical instabilities, especially disk tearing induced by differential Lense–Thirring precession in misaligned accretion disks, provide another possible mechanism for recurrent episodes of enhanced accretion \citep{raj_nixon2021}. Unlike the classical thermal-viscous instability, disk tearing can produce variability on dynamical or precessional timescales, potentially avoiding the problem of excessively long recurrence periods. However, the conditions required for tearing, its dependence on the accretion state, and the resulting spectral evolution remain insufficiently explored in the context of QPEs.

Orbital models, including repeated interactions of a stellar remnant or another orbiting body with the accretion disk, can naturally reproduce the characteristic recurrence times. Their main challenges concern the detailed spectral evolution, energetics, long-term period changes, and the relation between QPE activity and the luminosity state of the underlying accretion flow. Additional physical effects, including orbital precession, evolving disk structure, and interactions with the surrounding material, may substantially modify the simplest predictions.

An interesting example of variability potentially related to disk instabilities is provided by the CL AGN 1ES 1927+654. Following a major change in its accretion state, the source exhibited quasi-periodic X-ray variability with a timescale of approximately 12 days, interpreted by \citet{zhang2025} in terms of radiation-pressure instability during the rebuilding of the inner disk. This interpretation involves a transition region between the hot inner flow and the colder disk, with the evolution of a warm corona playing an important role. Although this variability differs from classical QPEs, it illustrates the potential importance of the evolving accretion flow geometry.
The evolution of warm corona was also observed in another CL AGN - Mrk 1018 \citep[see][]{noda2018}.

The most promising progress is likely to come from combining long-term monitoring with phase-resolved X-ray spectroscopy. The observed connections between QPE activity and the underlying luminosity state, as well as changes in recurrence periods, provide important constraints, but are not yet sufficient to distinguish between the proposed mechanisms. Temperature–luminosity hysteresis during individual eruptions, variability of absorption features and outflows, and correlated changes at other wavelengths offer additional tests. Their diagnostic value will depend on quantitative predictions from the competing models, including the evolution of the emitting area, temperature, spectral shape, and recurrence timescale.

\begin{svgraybox}
At present, no single model provides a fully satisfactory explanation of all observed QPE properties. Further theoretical development, together with systematic observations of a larger sample of sources over longer timescales, is essential to determine whether QPEs represent a common physical phenomenon or several different manifestations of accretion and orbital dynamics around massive black holes.
\end{svgraybox}

\begin{acknowledgement}
This work has received funding from the European Research Council (ERC) under
the European Union’s Horizon 2020 research and innovation program (grant
agreement No. [951549]). The Czech-Polish Mobility program of the two
Academies of Sciences, titled “Appearance and dynamics of accretion onto black
holes”, is greatly appreciated. AJ was supported by the grant 2023/50/A/ST9/00527 from Polish
National Science Center. We acknowledge computational support from the Interdisciplinary Center for Mathematical Modeling, ICM, of the Warsaw University, under allocation g104-2641, and the Polish high-performance computing infrastructure PLGrid (HPC Center: ACK Cyfronet AGH) for providing computer facilities and support within computational grant no. PLG/2025/018232.
BC acknowledges the OPUS-LAP/GA CR-LA
bilateral project (2021/43/I/ST9/01352/OPUS 22 and GF23-04053L).
MS acknowledges the Czech Science Foundation (GA\v{C}R) grant no. 26-23342I.
\end{acknowledgement}



\begin{thebibliography}{99}

\bibitem[Abramowicz et al.(1988)]{abramowicz1988}
M.~A.~Abramowicz, B.~Czerny, J.~P.~Lasota and E.~Szuszkiewicz,
Astrophys. J. \textbf{332}, 646 (1988).

\bibitem[Alush \& Stone(2025)]{Alush2025}
Y.~Alush and N.~C.~Stone,
Astrophys. J. \textbf{993}, 14 (2025).

\bibitem[Arcodia et al.(2021)]{arcodia2021}
R.~Arcodia, A.~Merloni, K.~Nandra \textit{et al.},
Nature \textbf{592}, 704 (2021).

\bibitem[Arcodia et al.(2022)]{arcodia2022}
R.~Arcodia, G.~Miniutti, G.~Ponti \textit{et al.},
Astron. Astrophys. \textbf{662}, A49 (2022).

\bibitem[Arcodia et al.(2024)]{arcodia2024}
R.~Arcodia, Z.~Liu, A.~Merloni \textit{et al.},
Astron. Astrophys. \textbf{684}, A64 (2024).

\bibitem[Arcodia et al.(2026)]{Arcodia2026}
R.~Arcodia, G.~Miniutti, J.~Chakraborty \textit{et al.},
Astrophys. J. \textbf{1003}, 148 (2026).

\bibitem[Asahina \& Ohsuga(2024)]{Asahina2024}
Y.~Asahina and K.~Ohsuga,
Astrophys. J. \textbf{973}, 45 (2024).

\bibitem[Balbus \& Hawley(1991)]{balbus1991}
S.~A.~Balbus and J.~F.~Hawley,
Astrophys. J. \textbf{376}, 214 (1991).

\bibitem[Baldini et al.(2026a)]{2026arXiv260709850B}
P.~Baldini, A.~Rau, A.~Merloni \textit{et al.},
arXiv:2607.09850 (2026a).

\bibitem[Baldini et al.(2026b)]{2026Baldini}
P.~Baldini, A.~Rau, A.~Merloni \textit{et al.},
Astron. Astrophys. \textbf{706}, L15 (2026b).

\bibitem[Ballantyne et al.(2024)]{ballantyne2024}
D.~R.~Ballantyne, V.~Sudhakar, D.~Fairfax \textit{et al.},
Mon. Not. R. Astron. Soc. \textbf{530}, 1603 (2024).

\bibitem[Begelman \& Pringle(2007)]{begelman2007}
M.~C.~Begelman and J.~E.~Pringle,
Mon. Not. R. Astron. Soc. \textbf{375}, 1070 (2007).

\bibitem[Begelman et al.(2015)]{2015begelman}
M.~C.~Begelman, P.~J.~Armitage and C.~S.~Reynolds,
Astrophys. J. \textbf{809}, 118 (2015).

\bibitem[Belloni et al.(1997)]{belloni1997}
T.~Belloni, M.~M{\'e}ndez, A.~R.~King, M.~van~der~Klis and J.~van~Paradijs,
Astrophys. J. Lett. \textbf{488}, L109 (1997).

\bibitem[Blaes et al.(2025)]{2025blaes}
O.~Blaes, Y.-F.~Jiang, J.-P.~Lasota and G.~Lipunova,
Space Sci. Rev. \textbf{221}, 120 (2025).

\bibitem[Chakraborty et al.(2025)]{Chakraborty2025}
J.~Chakraborty, E.~Kara, R.~Arcodia \textit{et al.},
Astrophys. J. Lett. \textbf{983}, L39 (2025).

\bibitem[Chakraborty et al.(2026a)]{Chakraborty2026}
J.~Chakraborty, E.~Kara, W.~Lu \textit{et al.},
Astrophys. J. Lett. \textbf{1010}, L37 (2026a).

\bibitem[Chakraborty et al.(2026b)]{2026Chakraborty}
J.~Chakraborty, S.~A.~Rappaport, R.~Arcodia \textit{et al.},
arXiv:2602.16776 (2026b).

\bibitem[Czerny \& Zycki(1994)]{czerny1994}
B.~Czerny and P.~T.~Zycki,
Astrophys. J. Lett. \textbf{431}, L5 (1994).

\bibitem[Czerny et al.(2003)]{czerny2003}
B.~Czerny, M.~Niko{\l}ajuk, A.~R{\'o}{\.z}a{\'n}ska \textit{et al.},
Astron. Astrophys. \textbf{412}, 317 (2003).

\bibitem[Czerny et al.(2009)]{czerny2009}
B.~Czerny, A.~Siemiginowska, A.~Janiuk, B.~Nikiel-Wroczy{\'n}ski and {\L}.~Stawarz,
Astrophys. J. \textbf{698}, 840 (2009).

\bibitem[Dewangan et al.(2000)]{dewangan2000}
G.~C.~Dewangan, K.~P.~Singh, Y.~D.~Mayya and G.~C.~Anupama,
Mon. Not. R. Astron. Soc. \textbf{318}, 309 (2000).

\bibitem[Dexter \& Begelman(2019)]{dexter2019}
J.~Dexter and M.~C.~Begelman,
Mon. Not. R. Astron. Soc. \textbf{483}, L17 (2019).

\bibitem[Fan et al.(2025)]{Fan2025}
N.~Fan, J.~F.~Steiner, C.~Bambi \textit{et al.},
Astrophys. J. \textbf{984}, 31 (2025).

\bibitem[Fragile et al.(2026)]{fragile2026}
P.~C.~Fragile, M.~J.~Middleton, B.~Brasseur, D.~A.~Bollimpalli and Z.~Smith,
Mon. Not. R. Astron. Soc. \textbf{548}, stag711 (2026).

\bibitem[Gianolli et al.(2023)]{Gianolli2023}
V.~E.~Gianolli, D.~E.~Kim, S.~Bianchi \textit{et al.},
Mon. Not. R. Astron. Soc. \textbf{523}, 4468 (2023).

\bibitem[Giustini et al.(2020)]{giustini2020}
M.~Giustini, G.~Miniutti and R.~D.~Saxton,
Astron. Astrophys. \textbf{636}, L2 (2020).

\bibitem[Giustini et al.(2024)]{giustini2024}
M.~Giustini, G.~Miniutti, R.~Arcodia \textit{et al.},
Astron. Astrophys. \textbf{692}, A15 (2024).

\bibitem[Gronkiewicz et al.(2023)]{gronkiewicz2023}
D.~Gronkiewicz, A.~R{\'o}{\.z}a{\'n}ska, P.-O.~Petrucci and R.~Belmont,
Astron. Astrophys. \textbf{675}, A198 (2023).

\bibitem[Grz{\k{e}}dzielski et al.(2017)]{grzedzielski2017}
M.~Grz{\k{e}}dzielski, A.~Janiuk, B.~Czerny and Q.~Wu,
Astron. Astrophys. \textbf{603}, A110 (2017).

\bibitem[Guolo et al.(2024)]{Guolo2024}
M.~Guolo, S.~Gezari, Y.~Yao \textit{et al.},
Astrophys. J. \textbf{966}, 160 (2024).

\bibitem[Guolo et al.(2025)]{Guolo_fitting_2025}
M.~Guolo, A.~Mummery, T.~Wevers \textit{et al.},
Astrophys. J. \textbf{985}, 146 (2025).

\bibitem[Hameury et al.(2000)]{2000Hameury}
J.-M.~Hameury, J.-P.~Lasota and B.~Warner,
Astron. Astrophys. \textbf{353}, 244 (2000).

\bibitem[Hern{\'a}ndez-Garc{\'\i}a et al.(2025a)]{2025NatAs...9..895H}
L.~Hern{\'a}ndez-Garc{\'\i}a, J.~Chakraborty, P.~S{\'a}nchez-S{\'a}ez \textit{et al.},
Nat. Astron. \textbf{9}, 895 (2025a).

\bibitem[Hern{\'a}ndez-Garc{\'\i}a et al.(2025b)]{2025hernandez}
L.~Hern{\'a}ndez-Garc{\'\i}a, P.~S{\'a}nchez-S{\'a}ez, J.~Chakraborty \textit{et al.},
Astron. Astrophys. \textbf{703}, A263 (2025b).

\bibitem[Ingram et al.(2023)]{Ingram2023}
A.~Ingram, M.~Ewing, A.~Marinucci \textit{et al.},
Mon. Not. R. Astron. Soc. \textbf{525}, 5437 (2023).

\bibitem[Janiuk(2020)]{2020janiuk}
A.~Janiuk,
in \textit{Multifrequency Behaviour of High Energy Cosmic Sources -- XIII} (Palermo, 3--8 June 2019), p.~48 (2020).

\bibitem[Janiuk \& Czerny(2007)]{2007A&A...466..793J}
A.~Janiuk and B.~Czerny,
Astron. Astrophys. \textbf{466}, 793 (2007).

\bibitem[Janiuk \& Czerny(2011)]{janiuk2011}
A.~Janiuk and B.~Czerny,
Mon. Not. R. Astron. Soc. \textbf{414}, 2186 (2011).

\bibitem[Janiuk \& Misra(2012)]{2012A&A...540A.114J}
A.~Janiuk and R.~Misra,
Astron. Astrophys. \textbf{540}, A114 (2012).

\bibitem[Janiuk et al.(2000)]{janiuk2000}
A.~Janiuk, B.~Czerny and A.~Siemiginowska,
Astrophys. J. Lett. \textbf{542}, L33 (2000).

\bibitem[Janiuk et al.(2002)]{janiuk2002}
A.~Janiuk, B.~Czerny and A.~Siemiginowska,
Astrophys. J. \textbf{576}, 908 (2002).

\bibitem[Janiuk et al.(2015)]{2015A&A...574A..92J}
A.~Janiuk, M.~Grzedzielski, F.~Capitanio and S.~Bianchi,
Astron. Astrophys. \textbf{574}, A92 (2015).

\bibitem[Jiang \& Pan(2025)]{Jiang2025}
N.~Jiang and Z.~Pan,
Astrophys. J. Lett. \textbf{983}, L18 (2025).

\bibitem[Jiang et al.(2013)]{2013jiang}
Y.-F.~Jiang, J.~M.~Stone and S.~W.~Davis,
Astrophys. J. \textbf{778}, 65 (2013).

\bibitem[Jiang et al.(2025)]{2025ApJ...988...43J}
Y.-F.~Jiang, O.~Blaes, I.~Kaul and L.~Zhang,
Astrophys. J. \textbf{988}, 43 (2025).

\bibitem[Kato et al.(1998)]{1998kato}
S.~Kato, J.~Fukue and S.~Mineshige (eds.),
\textit{Black-hole accretion disks} (Kyoto University Press, Kyoto, 1998).

\bibitem[Kaur et al.(2023)]{Kaur2023}
K.~Kaur, N.~C.~Stone and S.~Gilbaum,
Mon. Not. R. Astron. Soc. \textbf{524}, 1269 (2023).

\bibitem[Kosec et al.(2025)]{Kosec2025}
P.~Kosec, E.~Kara, L.~Brenneman \textit{et al.},
Astrophys. J. \textbf{978}, 10 (2025).

\bibitem[Lan{\v{c}}ov{\'a} et al.(2026)]{lancova2026}
D.~Lan{\v{c}}ov{\'a}, M.~Wielgus, M.~Abramowicz \textit{et al.},
arXiv:2603.17922 (2026).

\bibitem[Larwood et al.(1996)]{larwoodd1996}
J.~D.~Larwood, R.~P.~Nelson, J.~C.~B.~Papaloizou and C.~Terquem,
Mon. Not. R. Astron. Soc. \textbf{282}, 597 (1996).

\bibitem[Lasota(2002)]{2002lasota}
J.-P.~Lasota,
in \textit{The Physics of Cataclysmic Variables and Related Objects}, ed. B.~T.~G{\"a}nsicke, K.~Beuermann and K.~Reinsch,
ASP Conf. Ser. \textbf{261}, 397 (2002).

\bibitem[Lightman \& Eardley(1974)]{lightman1974}
A.~P.~Lightman and D.~M.~Eardley,
Astrophys. J. Lett. \textbf{187}, L1 (1974).

\bibitem[Linial \& Metzger(2024)]{2024ApJ...973..101L}
I.~Linial and B.~D.~Metzger,
Astrophys. J. \textbf{973}, 101 (2024).

\bibitem[Liska et al.(2023a)]{liska_corona2023}
M.~Liska, C.~Hesp, A.~Tchekhovskoy \textit{et al.},
Nature \textbf{101}, 102012 (2023a).

\bibitem[Liska et al.(2023b)]{liska2023}
M.~T.~P.~Liska, N.~Kaaz, G.~Musoke, A.~Tchekhovskoy and O.~Porth,
Astrophys. J. Lett. \textbf{944}, L48 (2023b).

\bibitem[Lu \& Quataert(2023)]{lu2023}
W.~Lu and E.~Quataert,
Mon. Not. R. Astron. Soc. \textbf{524}, 6247 (2023).

\bibitem[Merloni \& Nayakshin(2006)]{merloni2006}
A.~Merloni and S.~Nayakshin,
Mon. Not. R. Astron. Soc. \textbf{372}, 728 (2006).

\bibitem[Meyer \& Meyer-Hofmeister(1981)]{meyer1981}
F.~Meyer and E.~Meyer-Hofmeister,
Astron. Astrophys. \textbf{104}, L10 (1981).

\bibitem[Middleton \& Ingram(2015)]{middleton2015}
M.~J.~Middleton and A.~R.~Ingram,
Mon. Not. R. Astron. Soc. \textbf{446}, 1312 (2015).

\bibitem[Middleton et al.(2025)]{Middleton2025}
M.~Middleton, A.~G{\'u}rpide, T.~M.~Kwan \textit{et al.},
Mon. Not. R. Astron. Soc. \textbf{537}, 1688 (2025).

\bibitem[Miniutti et al.(2019)]{miniutti2019}
G.~Miniutti, R.~D.~Saxton, M.~Giustini \textit{et al.},
Nature \textbf{573}, 381 (2019).

\bibitem[Miniutti et al.(2023a)]{minitti_revival2023}
G.~Miniutti, M.~Giustini, R.~Arcodia \textit{et al.},
Astron. Astrophys. \textbf{674}, L1 (2023a).

\bibitem[Miniutti et al.(2023b)]{miniutti2023}
G.~Miniutti, M.~Giustini, R.~Arcodia \textit{et al.},
Astron. Astrophys. \textbf{670}, A93 (2023b).

\bibitem[Mishra et al.(2016)]{2016Mishra}
B.~Mishra, P.~C.~Fragile, L.~C.~Johnson and W.~Klu{\'z}niak,
Mon. Not. R. Astron. Soc. \textbf{463}, 3437 (2016).

\bibitem[Nayakshin et al.(2000)]{Nayakshin_2000}
S.~Nayakshin, S.~Rappaport and F.~Melia,
Astrophys. J. \textbf{535}, 798 (2000).

\bibitem[Nicholl et al.(2024)]{nicholl2024}
M.~Nicholl, D.~R.~Pasham, A.~Mummery \textit{et al.},
Nature \textbf{634}, 804 (2024).

\bibitem[Nixon \& King(2012)]{nixon_king2012}
C.~J.~Nixon and A.~R.~King,
Mon. Not. R. Astron. Soc. \textbf{421}, 1201 (2012).

\bibitem[Nixon \& King(2016)]{Nixon_review_2016}
C.~Nixon and A.~King,
in \textit{Astrophysical Black Holes}, ed. F.~Haardt, V.~Gorini, U.~Moschella, A.~Treves and M.~Colpi,
Lecture Notes in Physics \textbf{905} (Springer, Berlin), p.~45 (2016).

\bibitem[Nixon et al.(2012)]{nixon2012}
C.~Nixon, A.~King, D.~Price and J.~Frank,
Astrophys. J. Lett. \textbf{757}, L24 (2012).

\bibitem[Noda \& Done(2018)]{noda2018}
H.~Noda and C.~Done,
Mon. Not. R. Astron. Soc. \textbf{480}, 3898 (2018).

\bibitem[Palit et al.(2024)]{palit2024}
B.~Palit, A.~R{\'o}{\.z}a{\'n}ska, P.~O.~Petrucci \textit{et al.},
Astron. Astrophys. \textbf{690}, A308 (2024).

\bibitem[Pan et al.(2022)]{Pan22}
X.~Pan, S.-L.~Li, X.~Cao, G.~Miniutti and M.~Gu,
Astrophys. J. Lett. \textbf{928}, L18 (2022).

\bibitem[Pan et al.(2023)]{Pan23}
X.~Pan, S.-L.~Li and X.~Cao,
Astrophys. J. \textbf{952}, 32 (2023).

\bibitem[Pan et al.(2025)]{Pan25}
X.~Pan, S.-L.~Li, X.~Cao, B.~Liu and W.~Yuan,
Astrophys. J. \textbf{989}, 196 (2025).

\bibitem[Papaloizou \& Pringle(1983)]{papaloizou1983}
J.~C.~B.~Papaloizou and J.~E.~Pringle,
Mon. Not. R. Astron. Soc. \textbf{202}, 1181 (1983).

\bibitem[Petrucci et al.(2018)]{petrucci2018}
P.-O.~Petrucci, F.~Ursini, A.~De~Rosa \textit{et al.},
Astron. Astrophys. \textbf{611}, A59 (2018).

\bibitem[Petrucci et al.(2020)]{petrucci2020}
P.-O.~Petrucci, D.~Gronkiewicz, A.~Rozanska \textit{et al.},
Astron. Astrophys. \textbf{634}, A85 (2020).

\bibitem[Pringle(1981)]{pringle1981}
J.~E.~Pringle,
Annu. Rev. Astron. Astrophys. \textbf{19}, 137 (1981).

\bibitem[Pringle et al.(1973)]{pringle1973}
J.~E.~Pringle, M.~J.~Rees and A.~G.~Pacholczyk,
Astron. Astrophys. \textbf{29}, 179 (1973).

\bibitem[Pugliese \& Stuchl{\'\i}k(2015)]{stuchlik2015}
D.~Pugliese and Z.~Stuchl{\'\i}k,
Astrophys. J. Suppl. Ser. \textbf{221}, 25 (2015).

\bibitem[Quintin et al.(2025)]{quintin2025}
E.~Quintin, N.~Khan, N.~A.~Webb \textit{et al.},
arXiv:2511.22520 (2025).

\bibitem[Raj \& Nixon(2021)]{raj_nixon2021}
A.~Raj and C.~J.~Nixon,
Astrophys. J. \textbf{909}, 82 (2021).

\bibitem[Raj et al.(2021)]{rajetal2021}
A.~Raj, C.~J.~Nixon and S.~Do{\u{g}}an,
Astrophys. J. \textbf{909}, 81 (2021).

\bibitem[R{\'o}{\.z}a{\'n}ska et al.(2015)]{rozanska2015}
A.~R{\'o}{\.z}a{\'n}ska, J.~Malzac, R.~Belmont, B.~Czerny and P.-O.~Petrucci,
Astron. Astrophys. \textbf{580}, A77 (2015).

\bibitem[S{\k{a}}dowski(2016)]{2016sadowski}
A.~S{\k{a}}dowski,
Mon. Not. R. Astron. Soc. \textbf{459}, 4397 (2016).

\bibitem[Sakimoto \& Coroniti(1981)]{sakimoto1981}
P.~J.~Sakimoto and F.~V.~Coroniti,
Astrophys. J. \textbf{247}, 19 (1981).

\bibitem[Schawinski et al.(2015)]{schawinski2015}
K.~Schawinski, M.~Koss, S.~Berney and L.~F.~Sartori,
Mon. Not. R. Astron. Soc. \textbf{451}, 2517 (2015).

\bibitem[Shakura \& Sunyaev(1973)]{ss73}
N.~I.~Shakura and R.~A.~Sunyaev,
Astron. Astrophys. \textbf{24}, 337 (1973).

\bibitem[Shakura \& Sunyaev(1976)]{ss76}
N.~I.~Shakura and R.~A.~Sunyaev,
Mon. Not. R. Astron. Soc. \textbf{175}, 613 (1976).

\bibitem[Shu et al.(2017)]{shu2017}
X.~W.~Shu, T.~G.~Wang, N.~Jiang \textit{et al.},
Astrophys. J. \textbf{837}, 3 (2017).

\bibitem[Smak(1984)]{Smak1984}
J.~Smak,
Publ. Astron. Soc. Pac. \textbf{96}, 5 (1984).

\bibitem[Sniegowska et al.(2020)]{sniegowska2020}
M.~Sniegowska, B.~Czerny, E.~Bon and N.~Bon,
Astron. Astrophys. \textbf{641}, A167 (2020).

\bibitem[{\'S}niegowska et al.(2023)]{sniegowska2023}
M.~{\'S}niegowska, M.~Grz{\c{e}}dzielski, B.~Czerny and A.~Janiuk,
Astron. Astrophys. \textbf{672}, A19 (2023).

\bibitem[Speicher \& Blaes(2025)]{speicher2025}
J.~Speicher and O.~Blaes,
Astrophys. J. \textbf{980}, 143 (2025).

\bibitem[Sukov{\'a} et al.(2016)]{2016sukova}
P.~Sukov{\'a}, M.~Grzedzielski and A.~Janiuk,
Astron. Astrophys. \textbf{586}, A143 (2016).

\bibitem[Szuszkiewicz(1990)]{szuszkiewicz1990}
E.~Szuszkiewicz,
Mon. Not. R. Astron. Soc. \textbf{244}, 377 (1990).

\bibitem[Walter \& Fink(1993)]{walter1993}
R.~Walter and H.~H.~Fink,
Astron. Astrophys. \textbf{274}, 105 (1993).

\bibitem[Wevers et al.(2025)]{weavers2025}
T.~Wevers, M.~Guolo, S.~Lockwood \textit{et al.},
Astrophys. J. Lett. \textbf{980}, L1 (2025).

\bibitem[Wevers et al.(2026)]{2026arXiv260715359W}
T.~Wevers, J.~Chakraborty, E.~Quintin, M.~Zaja{\v{c}}ek and M.~Giustini,
arXiv:2607.15359 (2026).

\bibitem[Wijers \& Pringle(1999)]{wijers1999}
R.~A.~M.~J.~Wijers and J.~E.~Pringle,
Mon. Not. R. Astron. Soc. \textbf{308}, 207 (1999).

\bibitem[Zhang et al.(2025a)]{2025zhang}
L.~Zhang, J.~M.~Stone, P.~D.~Mullen \textit{et al.},
Astrophys. J. \textbf{995}, 26 (2025a).

\bibitem[Zhang et al.(2025b)]{zhang2025}
W.~Zhang, X.~Pan, M.~Liu \textit{et al.},
Astron. Astrophys. \textbf{704}, A187 (2025b).

\end{thebibliography}
\end{document}